\documentclass[twocolumn]{aastex631}

\usepackage{ upgreek }
\usepackage{amsmath}
\usepackage{comment}

\accepted{January 18, 2025}

\shorttitle{Radio polarization from stellar bowshocks}
\shortauthors{del Valle, Santos-Lima \& Pohl}

\begin{document}

\title{Radio polarization from runaway star bowshocks-I. The general case}

\correspondingauthor{M. V. del Valle}
\email{mvdelvalle@usp.br}

\author{M. V. del Valle}
\affil{Instituto de Astronomia, Geofísica e Ciências Atmosféricas, Universidade de São Paulo \\
              Rua do Matão, 1226 - Cidade Universitária - 05508-090 São Paulo-SP, Brasil}
 
 \author{R. Santos-Lima}
\affiliation{Instituto de Astronomia, Geofísica e Ciências Atmosféricas, Universidade de São Paulo \\
              Rua do Matão, 1226 - Cidade Universitária - 05508-090 São Paulo-SP, Brasil}

\author{M. Pohl}
\affil{Institute of Physics and Astronomy, University of Potsdam \\
14476 Potsdam, Germany}
\affiliation{Deutsches Elektronen-Synchrotron DESY \\ 
Platanenallee 6, 15738 Zeuthen, Germany}

\begin{abstract}
High velocity stars move through the interstellar medium with $V >$ 30\,km\,s$^{-1}$. 
When the star has powerful winds,  under the appropriate conditions, the interaction of the wind with the interstellar 
material produces a system of shocks. The outer shock, called the bowshock, perturbs the ambient medium, 
heating and compressing the gas.
The dust in the compressed bowshock cools, producing infrared radiation. This emission appears as extended 
coma-shape structures. The discovery of radio nonthermal emission from two stellar bowshock nebulae indicates 
that these sources might be accelerating electrons up to relativistic energies. The produced nonthermal 
radio emission is most probably synchrotron which has a high degree of polarization.
In this work we model the synchrotron 
emission of runaway massive star bowshocks aiming to produce synthetic radio emission and polarization maps
for two frequencies: 1.40 and 4.86 GHz. 
We model the interacting plasmas in a steady-state regime by means of magnetohydrodynamics simulations and we 
compute the injection and transport of the relativistic electrons in the diffusion approximation. We include 
in the model the most important depolarization effects. Our main conclusions are i) the effects of Faraday 
rotation within the source are important at the lowest frequency considered, 
ii) inferring the local magnetic field direction from polarization measurements only can be misleading, 
iii) thermal radio emission produced by ionized plasma within the bowshock structure and surroundings 
can surpass the polarized one
for the considered frequencies,
and iv) the contribution from the background electrons is minor.   
\end{abstract}


\graphicspath{{./}}

\section{Introduction}

Massive stars modify their surroundings through their intense ultra-violet (UV) radiation field, through violent mass-ejection episodes, or by the continuous injection of mechanical power from their powerful winds. These wind-interstellar medium (ISM) interactions produce shocks (see e.g. \citealt{2018ApJ...864...19D}). 
A particular case of stellar wind collision with the ISM comes from runaway massive stars. These are stars with high spatial velocities ($V_{\star} > 30$\,km\,s$^{-1}$) that have been expelled from their formation sites \citep[e.g.,][]{2000ApJ...544L.133H,2011MNRAS.410..190T}. In the interaction of the wind of an O or B star with the medium a bowshock is formed, in some cases detectable in the infrared (IR) as expanded coma-shaped sources \citep[e.g.,][]{1988ApJ...329L..93V,2010ApJ...710..549K,2011ApJ...734L..26V,2019MNRAS.486.4947K}. The dust accumulated { ahead of} the bowshock is heated by the stellar photons, cooling down and emitting the IR signal. There are of the order of $\sim$ 700 stellar bowshocks\footnote{ The shock itself is a very thin structure, however it is common to call the whole structure formed by the interaction of the wind with the medium the \emph{bowshock}.} candidates cataloged so far \citep{2012A&A...538A.108P,2015A&A...578A..45P, 2016ApJS..227...18K}. These bowshocks are also known as stellar bowshock nebulae (SBNe).   

A special case is the the bowshock of the massive runaway star BD +43$^{\circ}$3654, detected at radio wavelengths. Spectral index values indicate that the emission is of nonthermal nature: produced by relativistic electrons interacting with a magnetic field \citep{2010A&A...517L..10B}.This radiation encloses information of the magnetic field in the source and confirms the existence of a population of relativistic particles (electrons) in situ.

There are 8 massive runaway star bowshocks that present radio emission cataloged so far, in E-BOSS I and II \citep{2012A&A...538A.108P,2015A&A...578A..45P}. The nature of the emission (thermal or nonthermal) cannot be completely established for the moment. This is because observations at two (at least) radio frequencies are needed for the task, hence dedicated observations are required. The fact that these sources are nonthermal emitters cannot be ruled out, even in the presence of thermal radio emission, as the case of colliding-wind binaries (see Sect.\,\ref{sec:results}).
\citet{2022MNRAS.512.5374V} searched for radio emission in 15 bowshocks from E-BOSS I and II, using the recently released Rapid ASKAP\footnote{ASKAP, the Australian SKA Pathfinder is a precursor to the Square Kilometer Array.} Continuum Survey. They identified three confident counterparts, three plausible ones, and three inconclusive candidates. The authors investigated the underlying radio emission mechanism. They concluded that the radio emission in four SBNe is likely dominated by free-free radiation, with a possible nonthermal contribution and that in two cases the emission appears to be dominated by synchrotron. Recently, \citet{2022A&A...663A..80M} observed the bowshock of BD +43$^{\circ}$3654 and the Bubble Nebula (NGC 7635) associated with BD+60$^{\circ}$2522 \citep[e.g.,][]{2019A&A...625A...4G} at radio frequencies. Both bowshocks show nonthermal radio emission, adding a new source to this class, currently populated by two objects.

The presence of relativistic electrons in bowshocks of runaway stars implies that these systems  
can be 
able to accelerate particles (at least electrons) up to relativistic energies. The evidence of particle acceleration in systems formed by stellar winds interacting with the ISM is of great importance because in the last years the potential role of stellar winds in producing Galactic cosmic rays is increasingly becoming of special interest. The nonthermal electrons in SBNe can radiate at other wavelengths. Theoretical models predict nonthermal radiation produced mainly by inverse Compton (IC) scattering of IR photons, at X-rays and gamma rays \citep{2012A&A...543A..56D,2014A&A...563A..96D,2016A&A...588A..36P,2018A&A...617A..13D,2018ApJ...864...19D}. These predictions establish that the bowshocks of runaway stars are gamma-ray source candidates. 

Searches for the predicted emission at X-rays and gamma rays have been performed. \citet{2017MNRAS.471.4452D} concluded that a clear identification of nonthermal X-ray emission from massive runaway bowshocks requires  a sensitivity improvement of at least one order of magnitude \citep[see also][]{2019AJ....157..176B}. At gamma rays, \citet{2014A&A...565A..95S} search for emission in {\it Fermi} archive data of the 28 bowshocks listed in the E-BOSS catalog \citep{2012A&A...538A.108P}, no associated emission was found but upper limits were derived. Similar conclusions were obtained at very high energy by the H.E.S.S. collaboration \citep{2018A&A...612A..12H}. \citet{2018ApJ...861...32S} found a possible association of two runaway star bowshocks ($\lambda$ Cephei and LS 2355) with two unidentified {\it Fermi} gamma-ray sources from the third {\it Fermi} 3FGL  catalog \citep{2015ApJS..218...23A}.

Synchrotron emission is linearly polarized. In the absence of any depolarizing mechanism the expected degree of polarization radiation  is very high, up to $\sim$ 70\% \citep[e.g.,][]{1979rpa..book.....R}. Polarization in combination with other observables is a great tool for constraining physical properties of the emitting regions and for estimating the direction of the magnetic field component in the plane of the sky responsible for the emission \citep[e.g.,][]{2009ApJ...696.1864S}. The degree of synchrotron intrinsic polarization is very high, hence even a non-detection can provide information of the physical conditions in the emitter.   
Polarization from massive runaway star bowshocks have been detected associated with thermal emission. In the last years, \citet{benaglia2021} presented deep radio-polarimetric observations of the bowshock of BD+43$^{\circ}$3654. The data revealed no linear polarization in the bowshock above 0.5\%.

In this work we aim to model the synchrotron emission and corresponding polarization at radio wavelengths of a general massive runaway bowshock. In a separate paper we apply the model to BD+43$^{\circ}$3654 and discuss this particular case. The model is separated in 3 stages: 1) obtaining the steady state structure of the SBN and the macroscopic variables (density, temperature, magnetic field and velocity fields), 2) computing the transport of high-energy electrons in the obtained structure, and 3) synthesizing radio emission and polarization maps.

This work is organized 
in the following sequence. 
In Section\,\ref{sec:model} we describe our model and procedures. The results and discussion are presented in Sect.\,\ref{sec:results} and \ref{sec:discussion}. Finally, in Sect.\,\ref{sec:conclusions} we  present our conclusions.

\section{Model}\label{sec:model}

The interaction of the stellar wind colliding with the incoming interstellar material 
(considering the referential where the star is at rest) 
forms a system of two shocks. One of the shocks propagates trough the wind, the reverse shock (or termination shock). This is a non-radiative fast 
shock, with velocity of the order of the wind velocity (with respect to the wind material), 
i.e. thousands of kilometers per second. The forward shock 
compresses the ISM, with its velocity of the order of the star velocity, between 30 to 100 kilometers per second  (with respect to the ambient material) and it is usually radiative\footnote{ We follow the standard definition of radiative shock: a shock in which the structure of the density and temperature is affected by radiation from the shock-heated matter \citep[e.g.,][]{2005Ap&SS.298...49D}. The post-shock material may well be optically thin, which would eliminate the need to solve a radiative transport equation, and this is the situation that we envision.}. This shock is known as the bowshock\footnote{However, through all the paper we refer by bowshock to the whole structure produced by the wind and the ISM interaction.}. The two shocked materials are separated by a tangential discontinuity. 
{The} interstellar material swept by the star is compressed and heated by the ultra-violet (UV) stellar photon field\footnote{The material is also heated by the shock, however this effect on dust heating is sub-dominant, see e.g., \citet{2014MNRAS.444.2754M}.}. The dust present in the ISM cools down, producing the optically thin IR coma-shaped, parsec-size structures that are imaged by infrared satellites. { Our model does not calculate the emission produced by the dust}.     

In order to obtain an accurate prediction for the polarization of the synchrotron emission expected from the SBN we model the system in 3 stages enumerated in what follows.     
For the first stage we use a numerical model for solving the wind-ISM collision by means of three-dimensional (3D) 
magnetohydrodynamic (MHD) simulations. We simulate the interaction of the stellar wind with the ambient medium until  the system  reaches a steady state. In the next stage, we calculate the 
 two-dimensional (2D, axisymmetric, averaged around the star's velocity direction)
distribution of high-energy electrons in the system 
 with reduced dimensionality, i.e, considering an average of the MHD fields around the direction of the star's velocity
(the population accelerated in the system plus the one from the galactic background), using the  
MHD solution for describing the scenario: shock position, density, velocity  and magnetic fields. We solve the diffusion-advection-losses transport equation in the test-particle regime for the electrons. In the final stage, we compute the polarized synchrotron emissivity and the thermal free-free emissivity (Bremsstrahlung) at radio wavelengths, and integrate the Stokes parameters of the radiation along different lines of sight to produce synthetic observational maps 
(considering the full 3D structure of the MHD fields).

\begin{table}
\begin{center}
\begin{tabular}{lr}
\hline
Parameter & Value \\
\hline
$M_{\star}$ & 40$\,$M$_{\odot}$ \\
$R_{\star}$ & $10^{12}\,$cm \\
$T_{\star}$ & $4.25 \times 10^{4}\,$K \\
$B_{\star}$ & $200\,$G \\
$V_{\star}$ & $40\,$km$\,$s$^{-1}$ \\
$\dot{M}_{\rm w}$ & $7\times 10^{-7}\,$M$_{\odot}\,$yr$^{-1}$ \\
$V_{\rm w}$ & $2000\,$km$\,$s$^{-1}$ \\
$L_{\rm w}$ & $10^{36}\,$erg$\,$s$^{-1}$\\
$V_{\rm rot}$ & $100\,$km$\,$s$^{-1}$ \\
$n_{\rm ISM}$ & 0.57$\,$cm$^{-3}$ \\
$T_{\rm ISM}$ & 8000$\,$K \\
$B_{\rm ISM}$ & 5$\,\mu$G \\
\hline
\end{tabular}
\caption{Model parameters for the star 
($\star$), its wind ('w') and rotation ('r'), and ambient medium
('ISM'). 
 See the text for an explanation of each parameter.}
\label{tabla}
\end{center}
\end{table}

\subsection{Magnetohydrodynamic modeling}\label{sec:mhydro}

We use the PLUTO code \citep{2007ApJS..170..228M} to solve the ideal magnetohydrodynamic equations using a similar approach as in \citet{2014MNRAS.444.2754M,2017MNRAS.464.3229M} and \citet{2018ApJ...864...19D}. 
We employ a 3D domain in a referential where the star is at rest at the origin.
The system of equations solved are the following:
\begin{eqnarray}
\frac{\partial \rho}{\partial t} + \mathbf{\nabla} \cdot \left(\rho
		      \mathbf{v}\right)  & = & 0 \label{mass} \\
\frac{\partial \left( \rho \mathbf{v} \right)}{\partial t} + \nabla\cdot\left[ \rho \mathbf{v} 
	\mathbf{v} + \left( p + \frac{B^2}{8 \pi} \right) \mathbf{I} - \frac{ \mathbf{B B} }{4 \pi} \right] 
& = & 
0, \label{neutral_mom} \\
\frac{\partial e}{\partial t} + \mathbf{\nabla} \cdot \left[ \left(e +
		p + \frac{B^2}{8 \pi}\right)\mathbf{v} - \frac{\mathbf{B}(\mathbf{v}\cdot\mathbf{B})}{4 \pi} \right] & = &
{\Phi}(T,\rho)\\
\frac{\partial \mathbf{B}}{\partial t} - \mathbf{\nabla}\times \left( \mathbf{v}\times \mathbf{B} \right) & = & 0, \label{B_eqn} \\
								  \nabla\cdot\mathbf{B} & = & 0 \label{divB},
\end{eqnarray}
where ${\bf v}$,  ${\bf B}$, $\rho$ and $p$ are the fluid velocity, magnetic field, density and pressure, respectively; 
 $\mathbf{I}$ is the unit dyadic tensor;
$e = p/(\gamma_{\rm g} - 1) + \rho v^2/2 + B^2/8\pi$ is the total energy density with $\gamma_{\rm g} = 5/3$  the ratio of specific heats for a monoatomic ideal gas. $\Phi$ represents the radiative energy gains (heating) and losses (cooling)
mechanisms important for the ISM phase considered{,  assuming solar abundances}. { The cooling term includes
the cooling of hydrogen, helium, and metals \citep[from][]{wiersma2009}, also hydrogen recombination and collisionally excited forbidden
lines \citep{1997ApJS..109..517R}, while the heating term is due to recombination of hydrogen. For further details see \citet[][]{2014MNRAS.444.2754M} and references therein.}

We consider a fiducial  O-type star, assuming  typical parameters for the star, its wind and the ISM. These are listed in the Table\,\ref{tabla}.

The equations are solved using 
a cartesian grid.
The $z-$axis is in the direction of the stellar motion. The computational domain has dimensions  
$-30 \le x,y \le +30\,$pc
and  
$-20\le z \le 10$\,pc 
and  
resolution of 
$384 \times 384 \times 192$.

The wind is constantly launched in a region 
$r^{2} = x^{2}+y^{2}+z^{2} < 1\,{\rm pc}$ 
centered at the origin. Its density is given by $\rho_{\rm w} = \dot{M}_{\rm w}/(4{\pi}v_{\rm w}r^{2})$. 
The initial magnetic field configuration is described below.

For the ISM magnetic field we use a typical value of $B_{\rm ISM} = 5\,\mu$G. 
We consider the ISM field to be uniform, and its direction makes an angle $\alpha$ with respect to the $-z$ axis.
We ran simulations for 3 values of $\alpha$: $0^{\circ}$, $30^{\circ}$, and $90^{\circ}$.

For the magnetic field in the stellar wind we use the wind compression model with magnetic field  \citep[e.g.,][]{1963idp..book.....P,1982ApJ...253..188V,1998ApJ...505..910I}. This model is valid for weak magnetic fields, i.e. that do not partake in the dynamics of the wind, which is expected in  
most massive stars. Evidence shows that massive stars rotate at hundreds of kilometers per second,
hence the our model considers rotating massive stars.

We are interested in the  asymptotic solution at large distances from the star. In the absence of rotation the magnetic field would be purely radial, decaying as $r^{-2}$ from the star. Due to  rotation the field lines wrap around the star, and therefore an initially radial field in the surface becomes toroidal at large distances. 
For simplicity, we consider only the case where the rotation axis of the star is aligned with the direction $z$, i.e., aligned with the direction of the star's velocity with respect to the ISM.
In this limit the field 
components, 
for a spherical wind, are \citep[see][]{1998ApJ...505..910I,2013ApJ...765...19I}:
\begin{equation}\label{eq:bphi}
B_{\phi} = -B_{\star}\sin{\theta}\frac{R_{\star}V_{\rm rot}}{r\,V_{\rm w}},
\end{equation}
\begin{equation}\label{eq:br}
B_{r} = B_{\star}\left(\frac{R_{\star}}{r}\right)^{2},
\end{equation}
where $\theta = \arccos{(z / \sqrt{x^2 + y^2})}$ is the stellar colatitude.

This model is valid as long as the magnetic field is weak, rooted on the star, and governed by the 
flow\footnote{ This condition can be expressed as the Mach Alfvenic number, this is, the ratio between the wind speed and the Alfven speed, being much smaller than unity.}. 
The magnetic field is considered negligible for accelerating the wind when the Poynting flux distribution is negligible. 
This last condition can be expressed as \citep[see][]{1998ApJ...505..910I}: 
\begin{eqnarray}\label{eq:Blimit}
B_{\star} \ll && 2.5 {\rm kG}
\left( \frac{\dot{M}}{10^{-5} M_{\odot}\,{\rm yr}^{-1}}\right)^{1/2}
\left( \frac{V_{\rm w}}{10^{3}\,{\rm km}\,{\rm s}^{-1}}\right)^{3/2} \times \nonumber \\
&& \left( \frac{R_{\star}}{10\,R_{\odot}}\right)^{-1}
\left( \frac{V_{\rm rot}}{10^{2}\,{\rm km}\,{\rm s}^{-1}}\right)^{-1} .
\end{eqnarray}

\noindent For the fiducial model we use $B_{\star} \sim 200\,$G \citep[e.g.][]{2012SSRv..166..145W}, which is within the limit imposed by the above equation.

In the $z$  upper boundary the condition that fresh ISM 
material 
enters with ${\bf v} = - v_{\star} \,\hat{k}$ and ${\bf B}_{\rm ISM}$ is imposed.   We use {\sl outflow} conditions for all the other boundaries; also inflow into the domain is not allow to avoid boundary artifacts. 

The dynamic evolution is solved employing the second order Runge Kutta algorithm for the time integration.
Fluxes are computed using the Harten-Lax-van Leer solver, with parabolic spatial  reconstruction. 
The magnetic divergence is controlled using  the mixed hyperbolic/parabolic divergence cleaning technique of \citet{2002JCoPh.175..645D} implemented in PLUTO.

\begin{figure*}
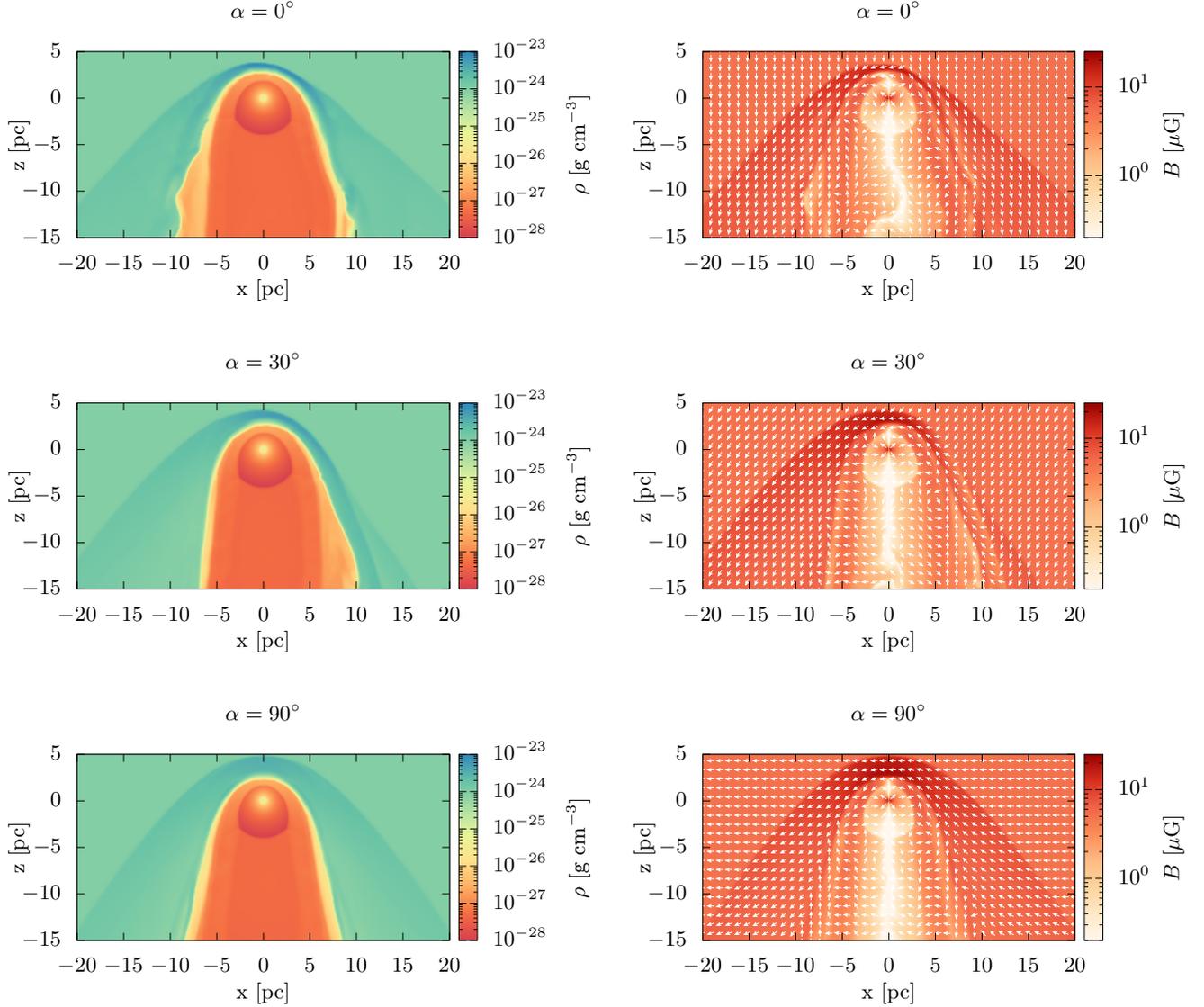

\begin{tabular}{c c}
	\input{./map_bowshock-3d-th00-maps_dens.tex} &
	\input{./map_bowshock-3d-th00-maps_magn.tex} \\
	\input{./map_bowshock-3d-th30-maps_dens.tex} &
	\input{./map_bowshock-3d-th30-maps_magn.tex} \\
	\input{./map_bowshock-3d-th90-maps_dens.tex} &
	\input{./map_bowshock-3d-th90-maps_magn.tex} \\
\end{tabular}
\caption{Maps of density (\emph{left}) and magnetic field magnitude (\emph{right}),  for the plane $y = 0$, corresponding to the 
``steady state'' solution of the collision of the wind of a massive star moving at $V_{\star} = 40$\,km\,s$^{-1}$ with the ISM. The white arrows in the right panel show the magnetic field direction. No turbulent magnetic field component is superimposed to the magnetic field in these maps (that is, they correspond to the models $\delta B_{\rm turb} / B_{\rm reg} = 0$ in Section~\ref{sec:results}; see also \S~\ref{sec:polcal}).}
\label{fig:densitymap}
\end{figure*}

Fig.\,\ref{fig:densitymap} shows the 
maps of density (left) and magnetic field (right), for a time when the system has reached a ``steady state'', 
which we observe after $t = 1.5$\,Myr (or $t = 2.5$\,Myr for one of our models, with $\alpha = 30^{\circ}$; see below). 
We show the 3 cases considered for the value of 
$\alpha$ 
(the angle between $\mathbf{B}_{\rm ISM}$ and $-\mathbf{V}_{\star}$): $0^{\circ}, 30^{\circ}$ and $90^{\circ}$.
{ It should be noted that the MHD fields oscillate in time, especially the structures in the wake behind the star.}  
We arbitrarily choose a representative snapshot of the system in this state and call this the ``steady state'' solution.
The white arrows in the right panel show the magnetic field direction in the plane, while the color represents the total field intensity
$| \mathbf{B(r)} |$.
A system of two shocks develops during the interaction. The reverse shock (also denominated termination shock) in the wind material (enclosing an egg-shaped surface around the origin) and the forward shock (bow-shaped) that compresses the ambient medium. 
The approximate values of the sonic Mach number $M_{\rm cs}$ and Alfvenic Mach number $M_{\rm A}$ for the 
{ forward} and { reverse} shock are: $M_{\rm cs} \approx M_{\rm A} \approx 4$ and $M_{\rm cs} \approx 2000$, $M_{\rm A} \approx 200$, 
respectively\footnote{ Observe that the relevant Mach number for the dynamics is the fast-magnetosonic Mach number ($M_{\rm fm}$) 
which be easily obtained from the sonic and the Alfvenic Mach numbers: $M_{\rm fm} = (M_{\rm cs}^{-2} + M_{\rm A}^{-2})^{-1/2}$.}.
The shocked ISM cools efficiently, compressing the material by a factor higher than 4, 
which would be expected for an strong shock in a monoatomic gas. 
The magnetic field at the reverse shock position is as high as 
a few $\mu$G. 
The highest values are reached near the star surface, see Equations (\ref{eq:bphi}) and (\ref{eq:br}),
and in the bowshock where the ambient magnetic field is amplified by compression, reaching values as high as  
{$\approx 20\,\mu$G}.
Differences can be observed in the structures for the different values of $\alpha$. These differences arise 
{ because only the 
component of $\mathbf{B}_{\rm ISM}$ 
lying in the shock front is compressed. 
Therefore if the direction of the ISM magnetic field changes (i.e. $\alpha$), so does the amplified component 
of $\mathbf{B}_{\rm ISM}$. As a result, different values of  
the downstream field strength are expected for different inclinations.} 
Detailed descriptions of the bowshock formed in runaway stars can be found in numerous works, the interested reader is referred to, e.g., \citet{2014MNRAS.444.2754M,2016A&A...586A.111S,2017MNRAS.464.3229M,2020MNRAS.493.4172S,2022MNRAS.516.3284L,2022A&A...663A..10B}.

\subsection{Transport of non-thermal electrons}\label{sec:particles}

We solve the transport of electrons in the bowshock of the massive star described in the previous subsection, using  the solution of the MHD simulations. 
In order to make the numerical calculations feasible with our transport code and the computational resources available, 
we reduce the spatial dimensionality of the system to 2D, with axisymmetry. We project the $B_x$ and $B_y$ components onto $B_R$ and $B_{\phi}$ and averaged over the azimuthal angle. We employ a cylindrical coordinate system $(R,z)$.

The diffusion-advection-losses equation for relativistic electrons that follows $N(t,\,E,\,\vec{r})$ $\equiv$ number of particles $/$ unit energy $\times$ unit volume, is: 

\begin{eqnarray}\label{eq:transport}
 \frac{\partial N(t,\,E,\,\vec{r})}{\partial t}
= & \nabla \left(D(t,\,E,\,\vec{r})\nabla N(t,\,E,\,\vec{r})\right) \nonumber\\ 
  & - \,\nabla \left( \vec{v}(t,\,\vec{r})N(t,\,E,\,\vec{r})\right) \nonumber\\ 
  & - \,\frac{\partial}{\partial E} \left(P(t,\,E,\,\vec{r})\,N(t,\,E,\,\vec{r})  \right)  \nonumber\\ 
  & + \,Q(t,\,E,\,\vec{r}),
\end{eqnarray}

\noindent where the rhs first term represents the diffusion in space with diffusion coefficient $D(t,\,E,\,\vec{r})$, followed by the advection term with $\vec{v}(t,\,\vec{r})$ the fluid velocity; the third term corresponds to radiative losses where $P(t,\,E,\,\vec{r})$ is the energy loss rate for a particle with energy $E$. Finally $Q(t,\,E,\,\vec{r})$ is the injection function, i.e. number of injected particles $/$ (unit energy $\times$ unit volume $\times$ unit time). 
In this problem the dominant terms of the above equation are the transport terms: advection and diffusion. 
The radiative loss term is kept only for completeness. 

We solve Eq.\,(\ref{eq:transport}) in a 3D grid $\equiv$ $(E,\,R,\,z)$ using our own modular code \citep[see,][]{2015MNRAS.448..207D,2018MNRAS.475.4298D,2018ApJ...864...19D}. We inject continuously a population of relativistic ($E > m_e c^{2}$) electrons  at the reverse shock position $(R_{\rm rs}, z_{\rm rs})$. 
The particles follow a power-law distribution in energy (in the interval $[m_e c^{2},E_{\rm max}]$) of index $\alpha = 2$
(that is, $Q \propto E^{- \alpha}$), 
as expected from a diffusive shock acceleration mechanism, normalized to a fraction 
($\zeta = 0.01$) of the available mechanical power of the system $L_{\rm w} = 0.5\,\dot{M}_{\rm w}V_{\rm w}^2$, 
giving an injection power in relativistic accelerated electrons 
$L_{e} \sim 10^{34}$\,erg\,s$^{-1}$.

We assume a {\sl Galactic-like} diffusion coefficient \citep[see a discussion on the matter in][]{2018ApJ...864...19D}:
$  D(E) = 10^{25}\left({E}/{10\,{\rm GeV}}\right)^{1/2}\,{\rm cm}^{2}\,{\rm s}^{-1}.$ 
 For simplicity, and in order to make the calculations feasible with our current transport code, the 
diffusion is assumed to be homogeneous and isotropic. 
However, a more realistic approach would be to consider a diffusion coefficient anisotropic with respect to the local magnetic field. 
However, the dominant transport process for relativistic particles at the energies we are interested in is advection (see \citealt{2018ApJ...864...19D}).

The maximum energy that electrons can reach in our model is limited by the size of the acceleration region. Given the characteristics of the system the acceleration process then should proceed in a region of size the order of $l$ $\sim$ $1$\,pc. Imposing the condition that the precursor size should be smaller than 1\,pc and assuming Bohm diffusion for the acceleration,  we obtain the maximum energy, i.e. $E < 3\,e\,B_{\rm shock}V_{\rm shock}l/c$. We use  $E_{\rm max} = 10$\,GeV.

We solve the Equation \ref{eq:transport} in a discrete grid $(E, R, z) \in [ 1 \, {\rm keV}, \, 50 \, {\rm GeV} ] \times [ 0, \, 30 \, {\rm pc} ] \times [ -20, \, 10 \, {\rm pc}]$, using the finite-volumes method. 
The energy grid is logarithmically spaced and the spatial grid is uniform. 
The grid resolution is $(L,M,K) = (128,256,256)$.

Initially we assume $N(0,\,E, \,R, \,z) \equiv 0$, i.e. no particles inside the domain.  
The energy boundary conditions are $N(t,\,E\,>\,E_{\rm max}, \,R, \,z) = 0$ and  
$N(t,\,E\,<\,E_{\rm min}, \,R, \,z) = 0$. The outer boundary condition for $R$ and the 
top and bottom boundary conditions for $z$  are assumed as outflow (zero normal derivative). We adopt axial symmetry at the $R$ inner boundary. 
The treatment of the Galactic background cosmic ray electrons is described in the Sect.\,\ref{sec:galactic_cre}.

\subsection{Radiative transfer}\label{sec:polcal}

The synchrotron polarized emissivity 
is calculated using the distribution in energy and space of the electron population obtained by solving Eq.(\ref{eq:transport}) and the 
3D
magnetic field solution 
from the MHD calculations
 (see Fig.\,\ref{fig:densitymap}). 
We calculate the emissivity coefficient 
distribution $j_\nu$ for two radio frequencies, $\nu_1 = 1.40$ and $\nu_2 = 4.86$\,GHz \citep[these are the frequencies of the radio detection made in][]{2010A&A...517L..10B} in the simulation domain, see \citet{1979rpa..book.....R} for the expressions. 
For obtaining the intensity $I_{\rm SYNC}$ on the sky plane we integrate $j_\nu$ along the line of sight, considering different inclinations of the bowshock structure with respect to the line-of-sight. 
All integrations are performed considering parallel light rays, which is a good approximation if the source has a small angular size.  For simplicity, no instrumental effects, such as beam size, were included in the intensity maps produced. { We consider different projections of the bowshock 3D structure onto the plane of the sky. This projection is 
quantified by the angle $\theta$ between the star motion and the line of sight.}

\begin{figure*}
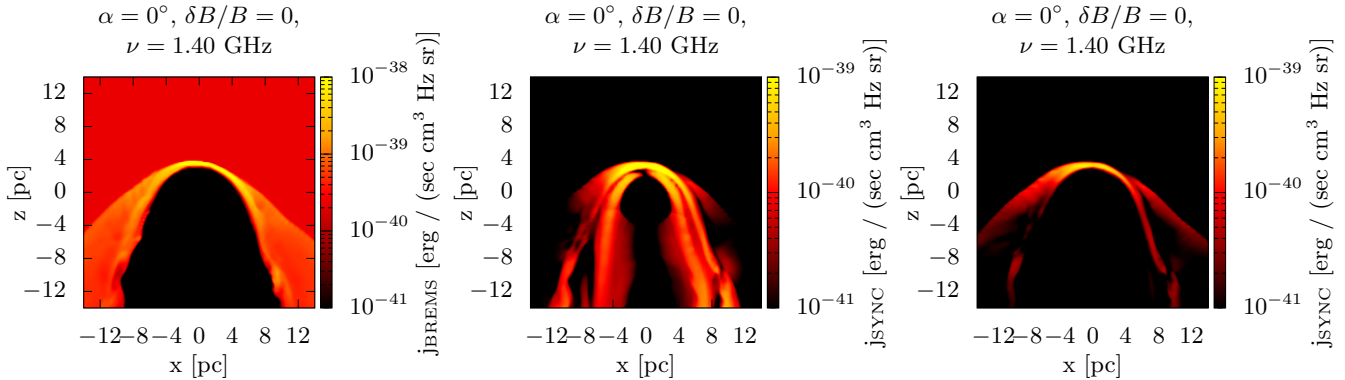

\begin{tabular}{c c c}
	\input{./map_brems_bowshock-3d-th00-ma0-emax10gev_1p40GHZ_CGS_smaller.tex} &
	\input{./map_sync_bowshock-3d-th00-ma0-emax10gev_1p40GHZ_CGS_smaller.tex} &
	\input{./map_sync_bowshock-3d-th00-ma0-bckg_1p40GHZ_CGS_smaller.tex}
\end{tabular}	
\caption{Emissivity distributions at $\nu = 1.40$\,GHz  for the plane $y = 0$, for our models with $\alpha = 0^{\circ}$ and $\delta B_{\rm turb} / B_{\rm reg} = 0$. \emph{Left:} Free-free (Bremsstrahlung) $j_{\rm BREMS}$ evaluated from the ``steady state'' MHD fields used with the transport equation (Eq.~\ref{eq:transport}; see also \S~\ref{sec:mhydro} and Fig.~\ref{fig:densitymap}).
\emph{Middle:} Total synchrotron emissivity calculated from the steady state solution of the cosmic ray electrons $N(E, \,R, \,z)$ injected at the reverse shock (see \S~\ref{sec:particles}).
\emph{Right:} Total synchrotron emissivity calculated from the steady state solution of the Galactic background of cosmic ray electrons $N_{\rm Gal}(E, \,R, \,z)$ (see \S~\ref{sec:galactic_cre}).}
\label{fig:emissivitymap}
\end{figure*}

The polarization is obtained by calculating the Stokes parameters $Q$ and $U$ (the polarization is linear, $V \equiv 0$) 
integrating
$j_{\nu \perp}$ and $j_{\nu \parallel}$, the polarized emission components. In the calculation of $Q$ and $U$ we take into account the effects of Faraday rotation in the position angle $\chi \equiv (1/2) \arctan (U/Q)$. This effect is important in wind-ISM systems \citep{2013ApJ...765...19I}. 
The polarized intensity is calculated along the optical paths via the integration of the Stokes parameters $Q$ and $U$:
\begin{gather}\label{eq:pol}
\frac{\rm d}{{\rm d}s}
 \begin{bmatrix}
  Q \\ U 
  \end{bmatrix}
 =
 (j_{\nu \perp} - j_{\nu \parallel})
  \begin{bmatrix}   
   \cos \left\{ 2 (\psi + \pi/2) \right\} \\ \sin \left\{ 2 (\psi + \pi/2) \right\} 
   \end{bmatrix}
   +
   2
 \begin{bmatrix}
  -U \\ Q 
  \end{bmatrix}
 \frac{{\rm d} \chi }{{\rm d}s}.
\end{gather}
\noindent where ${\rm d}s$ is the { element in the} optical path of integration, and $\psi (s)$ is the angle between the plane-of-sky magnetic field component at position $s$ and the reference direction for the position angle $\chi$. The degree of polarization at each point on the map is calculated by the ratio: $\Pi = {\sqrt{(Q^2+U^2)}}/{I_{\rm tot}}$, with $I_{\rm tot}$ the total intensity.

The magnetic field in the 
{ whole}
system is expected to have some degree of turbulence
due to natural instabilities and fluctuations which our resolution cannot resolve, and due to the natural fluctuations in the ISM 
that are also not represented
in our MHD simulations. 
Depolarization by turbulence can be 
{ important}
and should be included in the calculations. 
We model the turbulent component $\delta B_{\rm turb}$ superimposing to the regular field $B_{\rm reg}$ (given by the MHD simulation) an isotropic turbulent field which follows a 
Kolmogorov power-spectrum $\propto$ $k^{-5/3}$, with outer scale $k_{\rm inj}^{-1} = 1$\,pc. 
The degree of magnetic turbulence in the source is unknown, however we expect a high degree of turbulence. 
Here we assume 
a fiducial value 
$\delta B_{\rm turb} / B_{\rm reg} = 0.5$  (we also compute the case $\delta B_{\rm turb} / B_{\rm reg} = 1$, for comparison, see Sect.\,\ref{turbulence}), 
and we maintain $\delta B_{\rm turb}^2 + B_{\rm reg}^2 = B^2$ at each point of the domain, where $B$ is the local intensity of the 
magnetic field resulting from the MHD simulation. The optical path considered for the integration of the synchrotron emissivity is 
16\,pc, centered on the plane containing the star.

In order to obtain the degree of polarization we need to compute $I_{\rm tot}$. The competing radiation mechanism at these frequencies is thermal in nature: free-free emission, 
given that the UV radiation field from the star 
photo-ionizes 
the system surroundings, then $I_{\rm tot} = I_{\rm SYNC} + I_{\rm BREMS}$. The size of the Str\"omgren radius $R_{\rm S}$ is larger than the typical scale of a stellar bowshock \citep[e.g.,][]{2014MNRAS.444.2754M}; for a star with the parameters given in the Table\,\ref{tabla} it is $\sim$ 75\, pc. 
Not only the bowshock itself produces Bremsstrahlung emission, all the ionized material enclosing the Str\"omgren sphere projected in direction of the observer will contribute to $I_{\rm BREMS}$. We assume that the ISM properties inside the ionized region (and beyond a radius of 8\,pc centered on the star) contributing to the thermal radio emission are uniform. The standard formulas we use for estimating the free-free emission can be found in \citet{1979rpa..book.....R}.
The mean Gaunt factor for the free-free emission was taken from 
\citet{1998ppim.book.....S}. 
We also performed the calculation of the optical depth maps (due to the free-free absorption) and found it to be negligible at the studied frequencies.

Fig.~\ref{fig:emissivitymap} shows the free-free and synchrotron emissivity distributions at the frequency $\nu_1 = 1.40$\,GHz, for our models with $\delta B_{\rm turb} / B_{\rm reg} = 0$  and $\alpha = 0^{\circ}$. We show in different panels the synchrotron emissivity from the electrons accelerated in the reverse shock (middle panel) and the synchrotron emissivity from the Galactic cosmic ray electrons contribution (right panel; see \S~\ref{sec:galactic_cre}).
As described above, the optical path for integrating the radiation intensity (Stokes parameters) is taken arbitrarily from $-8$ to $+8\,$pc across a plane containing the star. This choice also reflects the limitation of the domain size in the MHD simulation and in the domain comprehended in the transport equation evolution. Although it can be seen that the strongest emissivity happens inside a radius of 8$\,$pc around the star, depending on the line-of-sight these limits of the optical path obviously truncates the contribution from the structures behind the star. As explained above, we 
sum the contribution from the Str\"omgren sphere outside these limits of the optical path ($-8$ and $+8\,$pc), which can underestimate the free-free intensity in the directions aligned with the shocked ISM around the contact discontinuity, and can 
overestimate the free-free intensity in the directions aligned with the low density plasma behind the star. At the same time, we are probably underestimating the synchrotron emissivity in the directions aligned  with the low density plasma behind the star. We expect this limitation to contribute with an error by a factor close to unity to our intensity maps presented in the next Section.

\section{Results}\label{sec:results}

We compute different projections of the bowshock 3D structure onto the plane of the sky. This projection is 
quantified by the angle $\theta$ between the star motion and the line of sight. We have computed the cases $\theta = 0, 30, 45$ and $90^{\circ}$. All the maps have resolution $128 \times 128$ and cover a region of $8 \times 8$ pc.

\subsection{Intensity}

\begin{figure*}
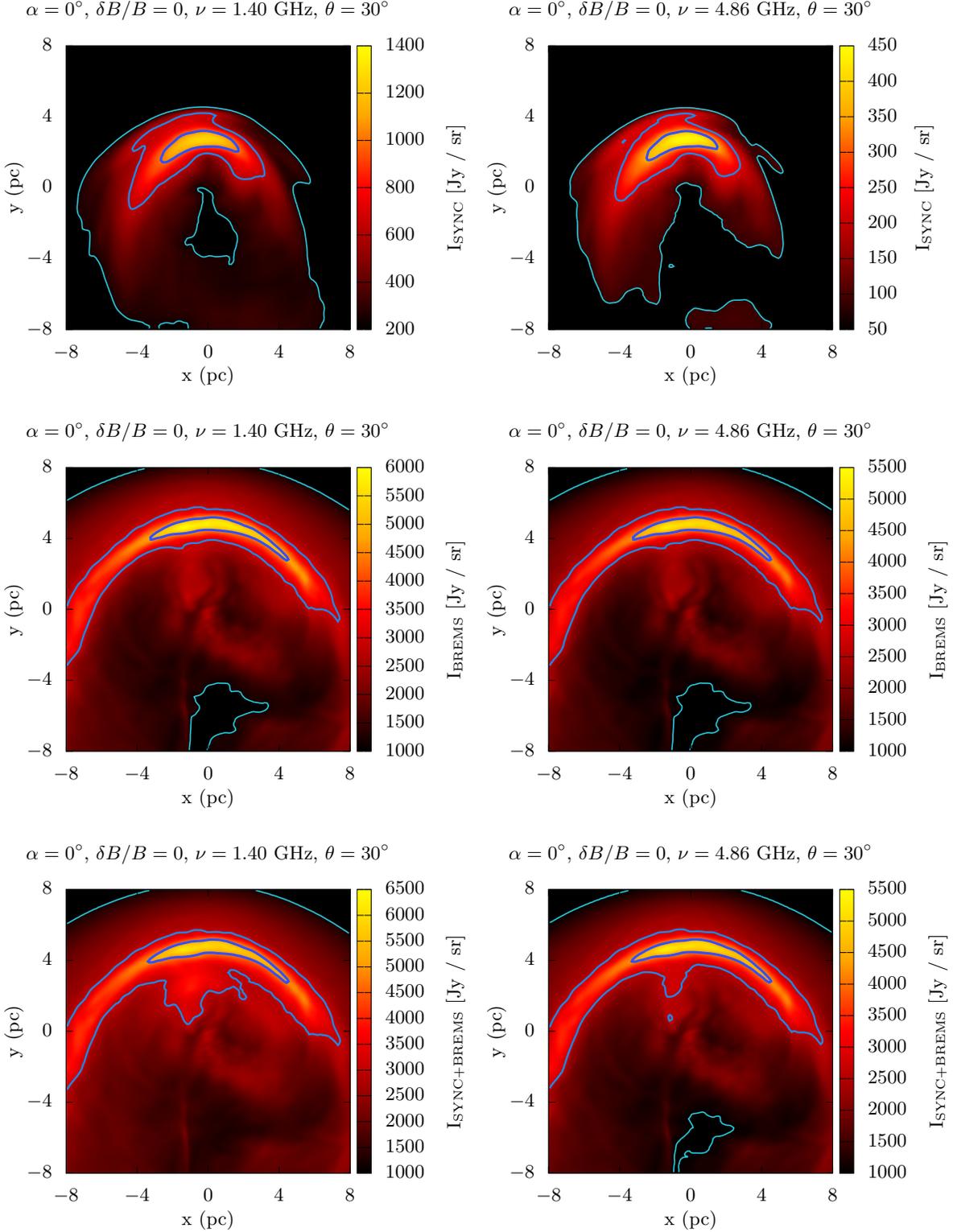

\begin{tabular}{c c}
	\input{./emis_sync_bowshock-3d-th00-ma0-emax10gev_1p40GHZ_CGS_0002.tex} &
	\input{./emis_sync_bowshock-3d-th00-ma0-emax10gev_4p86GHZ_CGS_0002.tex} \\
	\input{./emis_brems_bowshock-3d-th00-ma0-emax10gev_1p40GHZ_CGS_0002.tex} &
	\input{./emis_brems_bowshock-3d-th00-ma0-emax10gev_4p86GHZ_CGS_0002.tex} \\
	\input{./emis_sync_brems_bowshock-3d-th00-ma0-emax10gev_1p40GHZ_CGS_0002.tex} &
	\input{./emis_sync_brems_bowshock-3d-th00-ma0-emax10gev_4p86GHZ_CGS_0002.tex} \\
\end{tabular}
\caption{Specific intensity maps for the bowshock region projected onto the POS ($\theta = 30^{\circ}$ and $\alpha = 0^{\circ}$). The panels on the left correspond to $\nu$ = 1.4\,GHz, and those on the right to $\nu$ = 4.86\,GHz. The maps on the top correspond to synchrotron, the middle plots correspond to Bremsstrahlung, and the maps on the bottom show the sum of both contributions. The blue iso-intensity curves correspond to the values $I = 0.8$ (dark-blue), 0.5 (light-blue), and $0.2\,I_{\rm max}$ (cyan). See the text for further details.}
\label{fig:res1}
\end{figure*}

Fig.\,\ref{fig:res1} shows the intensity maps, for the 
fiducial
case $\theta = 30^{\circ}$  and $\alpha = 0^{\circ}$, at the two radio frequencies: 1.4 (left panels) and 4.86\,GHz (right panels). The upper plots show the synchrotron intensity maps, the middle plots show the Bremsstrahlung intensity maps, while the bottom plots show the sum of both contributions. 
Also plotted are contours of equal intensity, iso-intensity curves, with 
$I = 0.8$ (dark-blue), 0.5 (light-blue), and $0.2\,I_{\rm max}$ (cyan), 
where $I_{\rm max}$ is the maximum intensity value for each case. Observations have limiting lower intensities, imposed by the instrument sensitivity. In order to obtain a more realistic map here we adopt an arbitrary limit of $I = 0.2\,I_{\rm max}$.   

The synchrotron intensity grows towards lower frequencies, its value at 1.4\,GHz is more than twice that at 4.86\,GHz. 
In the case of the thermal emission the intensity practically remains within the same values. 
The nonthermal emission in this projection 
is stronger in the inner region of the bowshock structure, within a region of $\sim$ 4\,pc.
The thermal emission is much brighter than nonthermal radiation  in the shocked interstellar medium and provides the dominant contribution to the total radio flux.

\begin{figure}
\centering
\begin{tabular}{c}
	\input{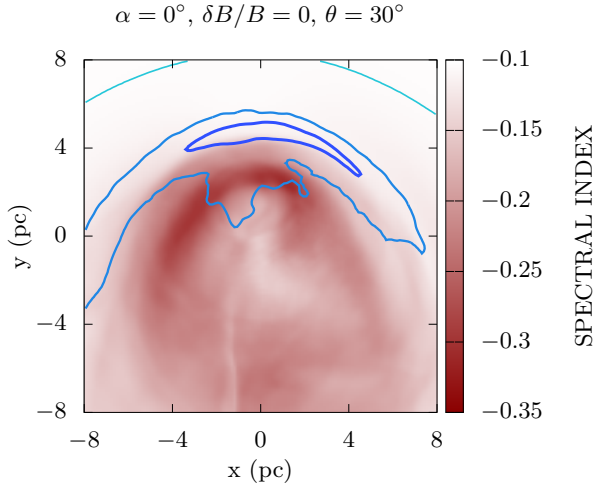}
\end{tabular}
\caption{Spectral index map for the bowshock region projected onto the POS ($\theta = 30^{\circ}$ and $\alpha = 0^{\circ}$). The blue iso-intensity curve corresponds to the values $I = 0.8$ (dark-blue), 0.5 (light-blue), and $0.2\,I_{\rm max}$ (cyan), for the total intensity at 1.40 GHz.}
\label{fig:res2}
\end{figure}

The contamination by thermal emission can be better appreciated in a spectral index map. 
Assuming that radio flux follows a power law in frequency $\nu^{s}$, we compute ${s}$ at each point of the map fitting a  power law between the obtained values at the two frequencies studied. The synchrotron emission, for a population of emitting electrons following a power-law distribution in energy $\propto$ $E^{-2}$, produces emissivity of spectral index $ s = -0.5$. 
In the case of Bremsstrahlung $s \approx - 0.1$ in the optically thin region. 
The map is shown in Fig.\,\ref{fig:res2}. The most negative values of the index, that corresponds to the nonthermal emission, are located in the inner \emph{bullet-shaped} region.
The iso-intensity curve of the total intensity acts as an observability limit, in a real observation only the region enclosed by the outer curve (cyan) would be visible in the map.  Within this region the values of the spectral index can reach values as negative as { $\sim -0.35$}. 

\subsection{Polarization degree} 

\begin{figure*}
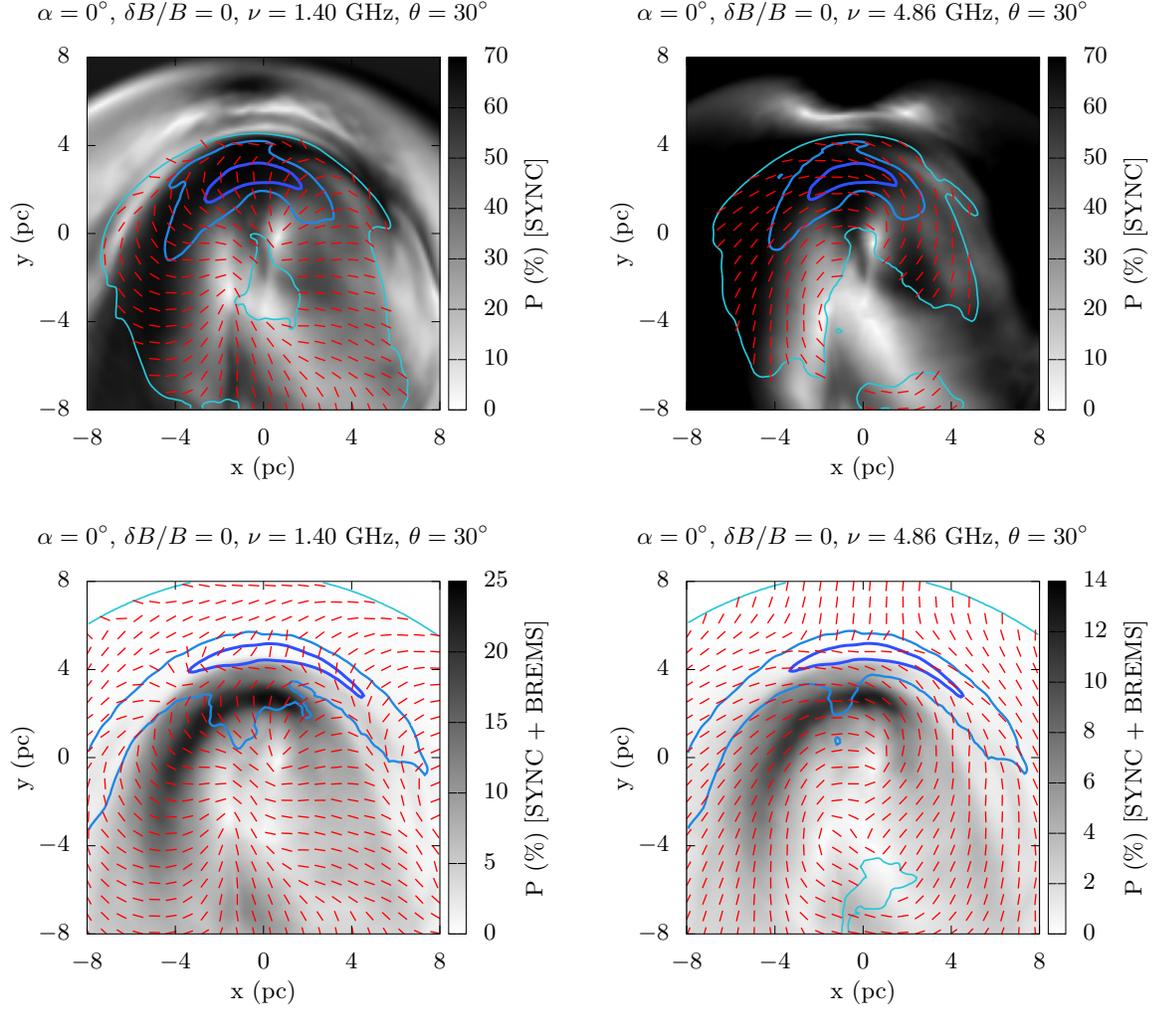

\begin{tabular}{c c}
	\input{./pol_sync_bowshock-3d-th00-ma0-emax10gev_1p40GHZ_CGS_0002.tex} &
	\input{./pol_sync_bowshock-3d-th00-ma0-emax10gev_4p86GHZ_CGS_0002.tex} \\
	\input{./pol_sync_brems_bowshock-3d-th00-ma0-emax10gev_1p40GHZ_CGS_0002.tex} &
	\input{./pol_sync_brems_bowshock-3d-th00-ma0-emax10gev_4p86GHZ_CGS_0002.tex} \\
\end{tabular}
\caption{Polarization degree maps for the bowshock region projected onto the POS ($\theta = 30^{\circ}$ and $\alpha = 0^{\circ}$). The panels on the left correspond to $\nu$ = 1.4\,GHz, and those on the right to $\nu$ = 4.86\,GHz. The maps on the top correspond to the polarization degree considering synchrotron emission only. The maps on the bottom panels show the polarization degree considering the total emission: thermal + nonthermal. The blue isocontours correspond to $I = 0.8$ (dark-blue), 0.5 (light-blue), and $0.2\,I_{\rm max}$ (cyan). Also shown in red is the position angle of the polarization (rotated by $90^{\circ}$). See the text for further details.}
\label{fig:res3}
\end{figure*}

The polarization degree map is shown in Fig.\,\ref{fig:res3}, for the frequencies  $\nu$ = 1.4\,GHz (left), and $\nu$ = 4.86\,GHz (right). 
The upper panels display the \emph{pure} polarization degree, i.e. the one expected 
solely from the synchrotron emission (including the Faraday rotation, see Eq.\,\ref{eq:pol}). 
The polarization percentage is very high, with regions reaching values as high as $P = 70\%$.
The most polarized regions correspond to the inner parts of the bowshock structure, close to the reverse shock in the shocked wind region. 
The bottom panels display the polarization degree considering the total intensity expected from the source at the frequencies of interest: 
synchrotron and Bremsstrahlung. Then $\Pi = {\sqrt{(Q^2+U^2)}}/{I_{\rm tot}}$, with $I_{\rm tot} = I_{\rm sync} + I_{\rm Brems}$. 
The presence of thermal emission decreases the degree of polarization to a maximum of { $25\%$} in the case of  $\nu = 1.4$\,GHz. 
The maps of Fig.\,\ref{fig:res3} also show, in red, the position angles of the polarized light (rotated by $90^{\circ}$) at each point.

The outer regions of the shown maps (upper panels) present high degrees of polarization. 
However, there is a tiny amount of synchrotron emission there (outside the limit of $I = 0.2\,I_{\rm max}$, delimited by the cyan contour). 
The most probable detection would come from the region enclosed by the blue curve 
($I = 0.5\,I_{\rm max}$, see caption of Fig.\,\ref{fig:res3}), 
where most of the flux is produced, 
and where the polarization degree is generally lower.

\subsection{Polarization direction}

\begin{figure*}
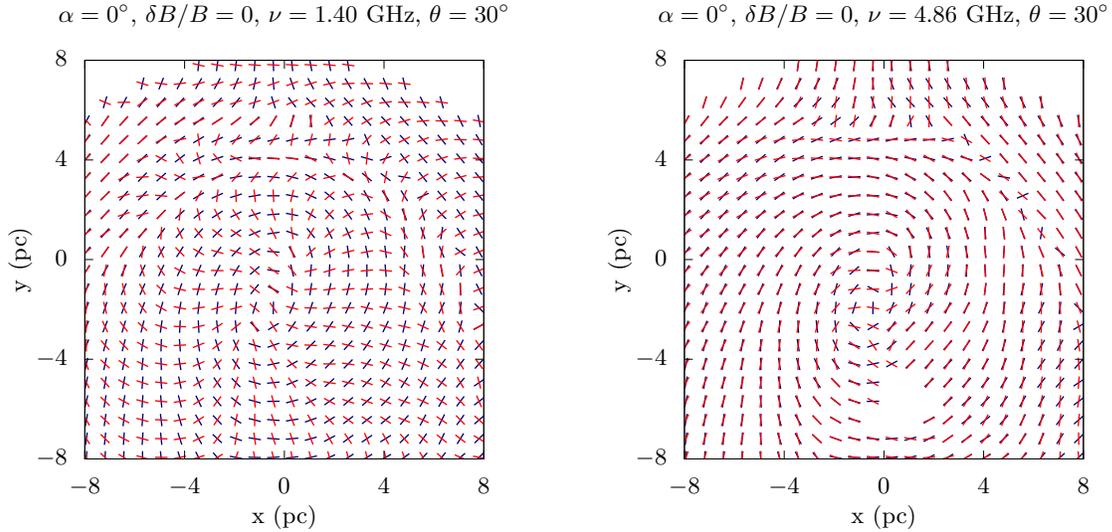

\begin{tabular}{c c}
	\input{./pa_sync_brems_bowshock-3d-th00-ma0-emax10gev_1p40GHZ_CGS_0002.tex} &
	\input{./pa_sync_brems_bowshock-3d-th00-ma0-emax10gev_4p86GHZ_CGS_0002.tex} \\
\end{tabular}
\caption{Polarization direction rotated 90$^{\circ}$ 
(red vectors)
for the bowshock region projected onto the POS ($\theta = 30^{\circ}$ and $\alpha = 0^{\circ}$). The blue vectors show the direction of the average magnetic field along the line of site. See the text for further details. Data is shown only inside regions where the total intensity $I$ is higher than $0.2\,I_{\rm max}$ (see isocontours in the bottom map of Fig.\,\ref{fig:res3}).}
\label{fig:res4}
\end{figure*}

Fig.\,\ref{fig:res4} shows the direction of polarization, rotated  90$^{\circ}$, expected from the synthetic maps. The effects of internal Faraday rotation can be seen in the differences appearing between the polarization direction at the two calculated frequencies. Faraday rotation is naturally stronger at lower frequencies, and the distortion in direction at $\nu = 1.4\,$GHz when compared to $\nu = 4.86\,$GHz is noticeable. Our results confirm that this effect is important and should be taken into account when analyzing radio-polarization from massive star winds.

The direction of polarization of the synchrotron emission 
is perpendicular to the local magnetic field. 
In Fig.\,\ref{fig:res4} we compare the directions of the 
polarization rotated 90$^{\circ}$ with the direction of the average magnetic field along the line of sight 
(inside the same optical path used for the calculation of the maps). 
Inside the region where most of the synchrotron emission is produced the directions do not coincide. 
This is not only due to Faraday rotation but due to the fact that the electrons are not uniformly distributed in the source. The regions of the strongest magnetic field might not be the regions of the strongest emission, because the density of relativistic particles can be low there.

\subsection{Effects of magnetic turbulence}\label{turbulence}

\begin{figure*}
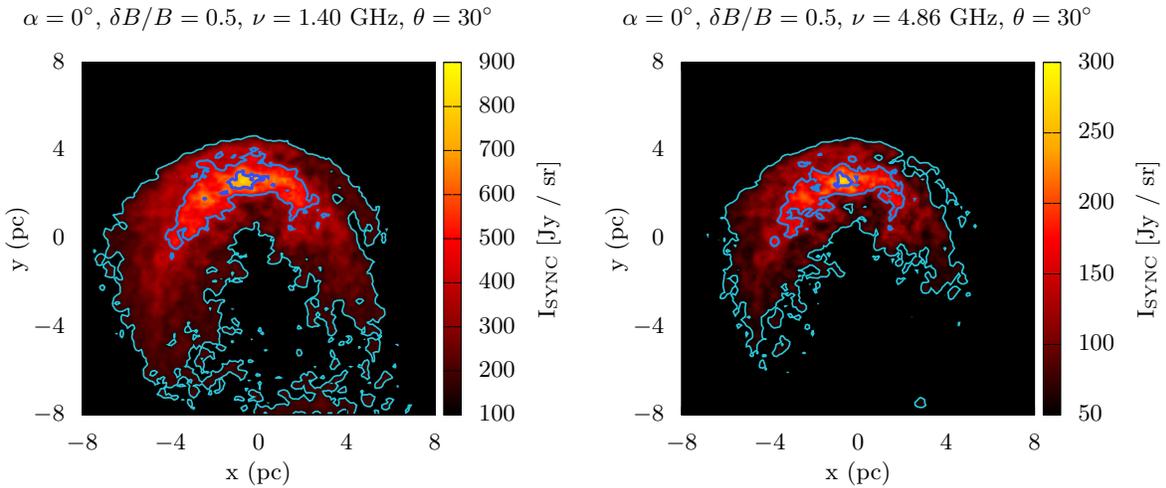

\begin{tabular}{c c}
	\input{./emis_sync_bowshock-3d-th00-ma0p5-emax10gev_1p40GHZ_CGS_0002.tex} &
	\input{./emis_sync_bowshock-3d-th00-ma0p5-emax10gev_4p86GHZ_CGS_0002.tex} \\
\end{tabular}
\caption{Intensity synchrotron maps for the bowshock region projected onto the POS ($\theta = 30^{\circ}$ and $\alpha = 0^{\circ}$) in the presence of a turbulent magnetic field component with ${\delta} B_{\rm turb}/B_{\rm reg} = 0.5$ . The panel on the left corresponds to $\nu$ = 1.4\,GHz, and that on the right to $\nu$ = 4.86\,GHz. The blue isocontours correspond to $I = 0.8$ (dark-blue), 0.5 (light-blue), and $0.2\,I_{\rm max}$ (cyan). See the text for further details.}
\label{fig:res5}
\end{figure*}

The effects of a magnetic turbulent component on the radio intensity can be seen in Fig.\,\ref{fig:res5}. 
This figure shows the nonthermal intensity maps, for $\theta = 30^{\circ}$, $\alpha = 0^{\circ}$, at frequencies: 1.4 (left) and 4.86\,GHz (right). 
The presence of turbulence 
(with amplitude $\delta B_{\rm turb} / B_{\rm reg} = 0.5$) 
modifies significantly the shape of the intensity structures. 
These synthetic maps resemble 
more the images at radio of the detected bowshocks. We do not consider here the inhomogeneities in density because we are 
concerned 
only with the depolarization produced by magnetic turbulence. However, density inhomogeneities expected to arise also from the action of turbulent motions would produce this nonuniform apparency in the thermal emission. 

The images at radio are expected to differ from the bowshock structures observed in the infrared, where the bowshock is imaged by the cooling dust, giving the appearance of diffuse emission in an intensity map.
\citep{1988ApJ...329L..93V,2016ApJS..227...18K}; for models see, e.g. \citet{2012A&A...548A.113D,2022A&A...663A..10B}.

\begin{figure*}
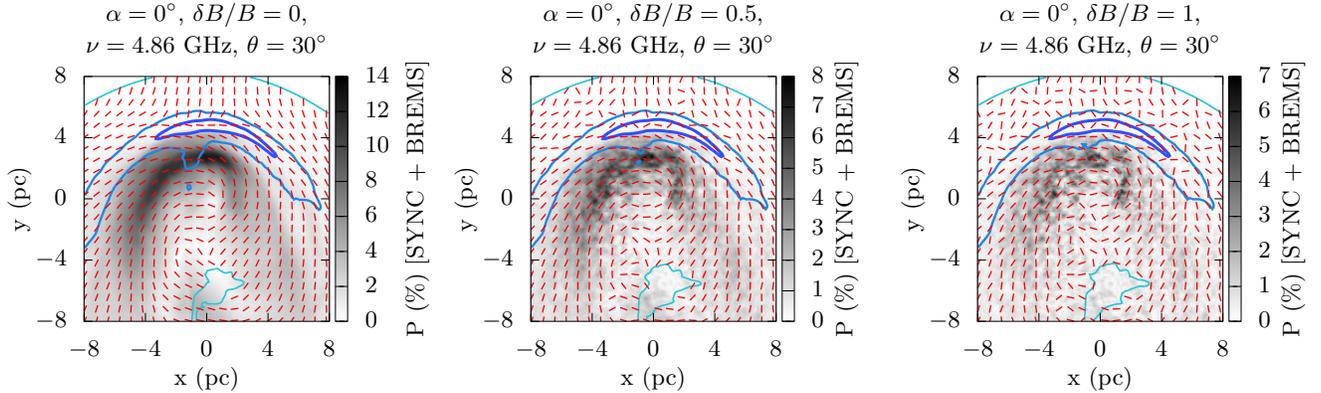

\begin{tabular}{c c c}
	\input{./pol_sync_brems_bowshock-3d-th00-ma0-emax10gev_4p86GHZ_CGS_smaller_0002.tex} &
	\input{./pol_sync_brems_bowshock-3d-th00-ma0p5-emax10gev_4p86GHZ_CGS_smaller_0002.tex} &
	\input{./pol_sync_brems_bowshock-3d-th00-ma1-emax10gev_4p86GHZ_CGS_smaller_0002.tex} \\
\end{tabular}
\caption{Polarization degree and position angle maps for the bowshock region projected onto the POS ($\theta = 30^{\circ}$  and $\alpha = 0^{\circ}$) at $\nu$ = 4.86\,GHz. 
The panels show the polarization degree considering the total emission (thermal + nonthermal) 
and a turbulent component of the magnetic field, with ${\delta} B_{\rm turb}/B_{\rm reg} = 0.0$ (left panel), 
$0.5$ (middle panel), and $1$ (right panel). 
The blue isocontours correspond to $I = 0.8$ (dark-blue), 0.5 (light-blue), and $0.2\,I_{\rm max}$ (cyan). See the text for further details.}
\label{fig:res6}
\end{figure*}

The effects of a turbulent magnetic field component on the polarization degree can be seen in Fig.\,\ref{fig:res6}, showing the degree of polarization at $\nu = 4.86$\,GHz. The polarization degree decreases significantly compared to the situation without this turbulent component, 
with peaks around { 8\%} when considering both thermal and synchrotron contributions and ${\delta} B_{\rm turb}/B_{\rm reg} = 0.5$ (see middle panel). 
For the case of a stronger turbulent magnetic field, with ${\delta} B_{\rm turb}/B_{\rm reg} = 1$, this effect is 
slightly higher, with the peak reducing to about 7\% (see right panel). 
The blue curves encloses the regions with $I = 0.8$ (dark-blue), 0.5 (light-blue), and $0.2\,I_{\rm max}$ (cyan). 


\subsection{Background electrons}\label{sec:galactic_cre}

The effect on the synchrotron emission produced by the electronic component of the Galactic cosmic rays should be taken into consideration. The background electrons would propagate into the bowshock, and in order to obtain their distribution in steady state we solve the transport equation (Eq.\,\ref{eq:transport}) in the same domain as for the injected electrons calculations. In this case, the injection is null, this is $Q(t,\,E,\,\vec{r}) = 0$. The outer boundary condition for $R$ and the inner and outer boundary conditions for $z$ 
{is}
the condition:
\begin{equation}
N(E) \equiv \frac{4\pi}{c} \,J_{\rm CR}(E),
\end{equation}   
where $J_{\rm CR}$ is isotropic Galactic cosmic ray flux. We use the parametrization from \citet{Potgieter_2015} \citep[see also,][]{2019ApJ...878...59B}, which was obtained using the flux measured by Voyager 1 in 2013.

\begin{figure}
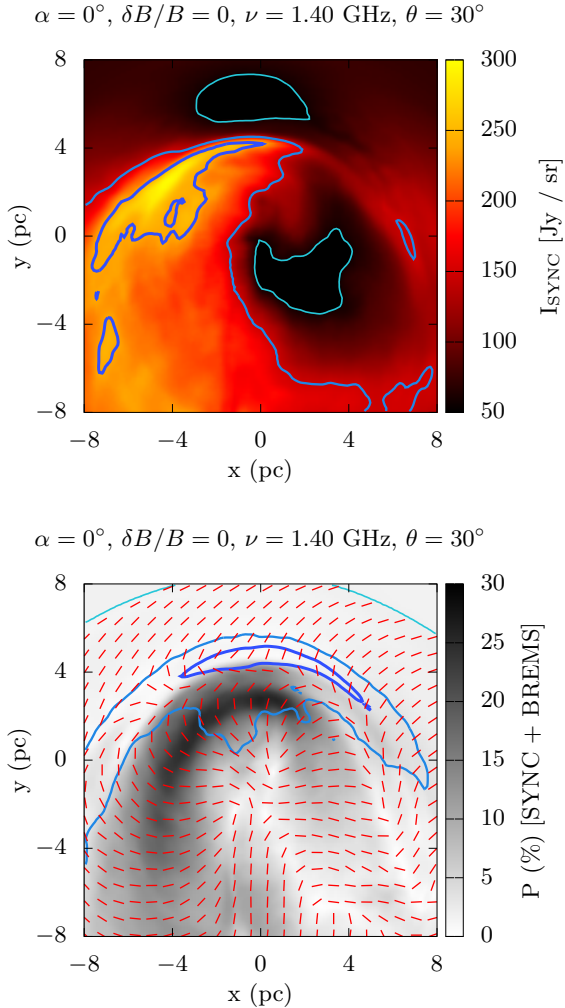

\centering
\begin{tabular}{c}
	\input{./emis_sync_bowshock-3d-th00-ma0-bckg_1p40GHZ_CGS_0002.tex} \\
	\input{./pol_sync_brems_bowshock-3d-th00-ma0-emax10gev-plus-bckg_1p40GHZ_CGS_0002.tex} \\
\end{tabular}
\caption{\emph{Top}: Specific intensity synchrotron map at $\nu = 1.4\,$GHz, for the bowshock region projected onto the POS ($\theta = 30^{\circ}$ and $\alpha = 0^{\circ}$) produced by the background cosmic ray electrons. \emph{Bottom}: Polarization degree map at $\nu = 1.4\,$GHz and position angle (rotated by $90^{\circ}$) for the total emission, considering the synchrotron emission from the injected electrons plus the background cosmic rays. The contours correspond to the values $I = 0.8$ (dark-blue), 0.5 (light-blue), and $0.2\,I_{\rm max}$ (cyan).}
\label{fig:background}
\end{figure}

Following an identical treatment as for the injected electrons, we compute the radio emissivity maps
(see the rightmost panel of Fig.\,\ref{fig:emissivitymap}), 
and we calculate the projection into the plane of the sky (POS). The resulted specific intensity synchrotron map at $\nu = 1.4\,$GHz is shown in the top plot of Fig.\,\ref{fig:background}. Comparing this result with the upper left plot shown in Fig.\,\ref{fig:res1} we see that the contribution from the background is lower than the one of the injected electrons. Also the maximum emissivity is reached at different regions. In the region where the background dominates (shocked ISM), both emissions are comparable. Another observation is that in this case the emission is highly asymmetric, this is because the geometrical effect of the radial component of the magnetic field in the wind to the total field: in one side it contributes positively to the total field (i.e., taking into account the ISM field) and negatively in the other.

The effects on the total polarization degree can be seen on the bottom plot of Fig.\,\ref{fig:background}. This map shows the total polarization percentage considering the contribution from the nonthermal injected electrons plus the 
Galactic background 
cosmic-ray component. 
The polarization 
keeps similar { degree} and 
spatial distribution  (see bottom left panel in Fig.\,\ref{fig:res3} for comparison). 

\begin{figure*}
\begin{tabular}{c c c}
	\input{./emis_sync_brems_bowshock-3d-th00-ma0p5-emax10gev-plus-bckg_1p40GHZ_CGS_smaller_0001.tex} &
	\input{./emis_sync_brems_bowshock-3d-th00-ma0p5-emax10gev-plus-bckg_1p40GHZ_CGS_smaller_0002.tex} &
	\input{./emis_sync_brems_bowshock-3d-th00-ma0p5-emax10gev-plus-bckg_1p40GHZ_CGS_smaller_0004.tex} \\
	\input{./emis_sync_brems_bowshock-3d-th30-ma0p5-emax10gev-plus-bckg_1p40GHZ_CGS_smaller_0001.tex} &
	\input{./emis_sync_brems_bowshock-3d-th30-ma0p5-emax10gev-plus-bckg_1p40GHZ_CGS_smaller_0002.tex} &
	\input{./emis_sync_brems_bowshock-3d-th30-ma0p5-emax10gev-plus-bckg_1p40GHZ_CGS_smaller_0004.tex} \\
	\input{./emis_sync_brems_bowshock-3d-th90-ma0p5-emax10gev-plus-bckg_1p40GHZ_CGS_smaller_0001.tex} &
	\input{./emis_sync_brems_bowshock-3d-th90-ma0p5-emax10gev-plus-bckg_1p40GHZ_CGS_smaller_0002.tex} &
	\input{./emis_sync_brems_bowshock-3d-th90-ma0p5-emax10gev-plus-bckg_1p40GHZ_CGS_smaller_0004.tex} \\
\end{tabular}
\caption{Intensity maps at $\nu = 1.4\,$GHz for the total emission (the contribution from the cosmic rays is also considered), for 3 projection angles $\theta = 0,\,30$ and $90^{\circ}$ 
(left, middle, and right column, respectively)
and for 3 values of $\alpha$ (angle between $\mathbf{B}_{\rm ISM}$ and $-\mathbf{V}_{\star}$): $0^{\circ}$ (upper panels), $30^{\circ}$ (middle panels), and $90^{\circ}$ (bottom panels). 
The contours correspond to the values $I = 0.8$ (dark-blue), 0.5 (light-blue), and $0.2\,I_{\rm max}$ (cyan).}
\label{fig:inte-angles}
\end{figure*}

\begin{figure*}
\begin{tabular}{c c c}
	\input{./pol_sync_brems_bowshock-3d-th00-ma0p5-emax10gev-plus-bckg_1p40GHZ_CGS_smaller_0001.tex} &
	\input{./pol_sync_brems_bowshock-3d-th00-ma0p5-emax10gev-plus-bckg_1p40GHZ_CGS_smaller_0002.tex} &
	\input{./pol_sync_brems_bowshock-3d-th00-ma0p5-emax10gev-plus-bckg_1p40GHZ_CGS_smaller_0004.tex} \\
	\input{./pol_sync_brems_bowshock-3d-th30-ma0p5-emax10gev-plus-bckg_1p40GHZ_CGS_smaller_0001.tex} &
	\input{./pol_sync_brems_bowshock-3d-th30-ma0p5-emax10gev-plus-bckg_1p40GHZ_CGS_smaller_0002.tex} &
	\input{./pol_sync_brems_bowshock-3d-th30-ma0p5-emax10gev-plus-bckg_1p40GHZ_CGS_smaller_0004.tex} \\
	\input{./pol_sync_brems_bowshock-3d-th90-ma0p5-emax10gev-plus-bckg_1p40GHZ_CGS_smaller_0001.tex} &
	\input{./pol_sync_brems_bowshock-3d-th90-ma0p5-emax10gev-plus-bckg_1p40GHZ_CGS_smaller_0002.tex} &
	\input{./pol_sync_brems_bowshock-3d-th90-ma0p5-emax10gev-plus-bckg_1p40GHZ_CGS_smaller_0004.tex} \\
\end{tabular}
\caption{Polarization degree maps at $\nu = 1.4\,$GHz for the total emission (the contribution from the cosmic rays is also considered), for 3 projection angles $\theta = 0,\,30$ and $90^{\circ}$ 
(left, middle, and right column, respectively)
and for 3 values of $\alpha$ (angle between $\mathbf{B}_{\rm ISM}$ and $-\mathbf{V}_{\star}$): $0^{\circ}$ (upper panels), $30^{\circ}$ (middle panels), and $90^{\circ}$ (bottom panels). 
The contours correspond to the values $I = 0.8$ (dark-blue), 0.5 (light-blue), and $0.2\,I_{\rm max}$ (cyan).}
\label{fig:angles}
\end{figure*}

The dependence of the total intensity with the viewing angle can be seen in 
the upper panels of 
Fig.\,\ref{fig:inte-angles} for $\nu = 1.4$\,GHz, including the intensity produced by the background electrons. 
We show maps for $\theta = 0,\,30$ and $90^{\circ}$ 
(left, center, and right column, respectively).
The case $\alpha = 0^{\circ}$ is shown in the top panels. 
We also show the dependence of the polarization degree with the viewing angle in Fig.\,\ref{fig:angles}
(top panels for $\alpha = 0^{\circ}$), 
where we show the final polarization degree, at the same frequency, again for $\theta = 0,\,30$ and $90^{\circ}$ 
(left, center, and right column, respectively). 
The percentage of polarized light is of the same order in all cases, but for $\theta = 90^{\circ}$, the inner polarization is higher.
However in these extended sources the polarization is not uniformly distributed in the domain. 
In all cases, a polarization degree of $\lessapprox 30\%$, is encircled inside the $I = 0.2\,I_{\rm max}$ contour. However, the polarization expected inside the $I = 0.8\,I_{\rm max}$ contour is much lower, of the order of $8-10\%$.

\subsection{Dependence on $\alpha$}

{
The angle between the ambient magnetic field vector and the star velocity direction, $\alpha$, might be important in the total polarization degree. As we have already discussed, the relative direction of $\mathbf{B}_{\rm ISM}$ slightly modifies the overall structure of the bow shock (see Fig.\,\ref{fig:densitymap}). The effects of $\alpha$ on the polarization can be appreciated in Fig.\,\ref{fig:angles}. 
The maps 
show the total polarization degree maps, at $\nu = 1.4\,$GHz for $\alpha = 0^{\circ}$, $30^{\circ}$ and 90$^{\circ}$ (top, middle and bottom panels, respectively). 
The projection angles are $\theta = 0^{\circ}$ (left column), $30^{\circ}$ (middle column), and $90^{\circ}$ (right column). 
Only the component of $\mathbf{B}_{\rm ISM}$ perpendicular to the shock normal is compressed, hence the cases with $\alpha \neq 0$ would present locally higher values of ${B}_{\rm ISM}$. 
Furthermore, in these cases the layer of shocked interstellar medium is wider, and therefore the integrated synchrotron emission is higher; 
this is evident in the contours of synchrotron intensity shown in the plots { (see also Fig.\,\ref{fig:inte-angles})}. In the extreme case of $\alpha = 90^{\circ}$, the final polarization degree can reach values { as high as $45\%$}. 
Also, for the oblique line-of-sight ($\theta = 30^{\circ}$) the regions of higher intensity { enclose} regions of high polarization, 
making this case the best case scenario for a polarization detection. { The side-on case (i.e. $\theta = 90^{\circ}$) with $\alpha = 90^{\circ}$ presents the higher polarization inside the regions of maximum intensity.}

}

\section{Discussion}\label{sec:discussion}

The results presented here are valid under the 
described 
conditions and assuming that the relativistic electrons are accelerated in the termination shock. A different acceleration scenario would produce a different result, especially the localization of the nonthermal emission.  
The map of polarization degree might also change if the nonthermal emission and the Bremsstrahlung are produced in the same regions of the source.  For example, in \citet{2022A&A...663A..80M} the authors proposed that the electrons are accelerated at the forward shock, hence in this scenario a lower polarization degree is expected.    

The maximum polarization degree (also considering the turbulent component of the magnetic field) 
is about 45\% in very localized spots. This corresponds to the case of $\alpha = 90^{\circ}$ and $\theta = 90^{\circ}$, with ${\delta} B_{\rm turb}/B_{\rm reg} = 0.5$.  However, a most probable case would be between these inclinations, a maximum value of  30\% would be observed inside the $0.2\,I_{\rm max}$ limit, and a most feasible value of 8-10\% inside $0.8\,I_{\rm max}$ has the greatest chance of been observed. 

The polarization degree can be much
lower than the one obtained here in a denser interstellar medium, where the thermal emission component is expected to be stronger. Detecting the IR structure requires high values of $n_{\rm ISM}$ \citep{2016ApJS..227...18K}, but detecting synchrotron radio  polarization is expected in faint bowshocks. 
However, a denser medium, like the one expected in the surroundings of BD+43$^{\circ}$3654, produces a smaller bowshock structure, where the reverse shock is closer to the star and therefore the magnetic field is stronger, increasing the synchrotron signal. 


Stronger winds produce more powerful shocks, however the bowshock structure changes and 
direct predictions cannot be made {\it a priori}. Because of the complexity of the system, we need to perform a dedicated study of the special case of BD+43$^{\circ}$3654, to be presented in a follow-up paper. 

We can conclude that the detection of radio emission from a runaway massive star bowshock makes the source a particle-accelerator candidate, a \emph{Particle Accelerating Stellar bowshock Nebula (PASBN)}. 
We cannot {\it a priori} reject a scenario where
particles are indeed being accelerated up to relativistic energies, even
in objects that are not classified as nonthermal radio emitters. Just like in the case of colliding-wind binaries the nonthermal emission produced by the locally accelerated particles might be undetectable under the thermal emission produced in the ionized bowshock itself and its surroundings. So far we have two confirmed objects of the class PASBN and 3 candidates from \citet{2022MNRAS.512.5374V}.\\

\section{Conclusions}\label{sec:conclusions}

In this paper we estimate the radio ($\sim$ GHz) polarization in the bowshock of a runaway massive star. We consider a standard O9 star with $\dot{M} = 7\times 10^{-7}$\,M$_{\odot}$\,yr$^{-1}$ and $V_{\rm wind} = 2000$\,km\,s$^{-1}$, moving at $40\,$km\,s$^{-1}$ in a medium with $n_{\rm ISM} = 0.57$\,cm$^{-3}$ (see Table\,\ref{tabla}). 
First, we obtained the 3D 
MHD ``steady-state'' structure formed by the collision of the stellar wind with the incoming interstellar medium (in the star's reference frame). Second, we compute the transport of relativistic electrons injected with a power-law in energy $\propto$ $E^{-2}$ at the position of the reverse shock, using the ``steady-state'' solution 
with reduced dimensionality (fields averaged around the axis parallel to the direction of the star's velocity)
of the previous step as the physical scenario for the particles. Finally, we calculate the synchrotron emission, thermal Bremsstrahlung and polarization degree at low radio frequencies and produce synthetic maps. We take into account the effects of a turbulent component of the magnetic field (with outer scale of 1 pc) and of the Galactic cosmic ray electrons. We arrive at the following conclusions:

\begin{enumerate}
\item The shape and maximum intensity of the bowshock at radio frequencies depend on the inclination of the structure with respect to the { plane of the sky}; 
\item The nonthermal emission is produced in the inner regions of the bowshock (near the reverse shock), while the thermal emission is mainly produced in the shocked interstellar medium;
\item Internal Faraday rotation of the polarization direction is not negligible for frequencies $\lesssim 1.4$ GHz. The polarization direction alone, at these frequencies, is not a good tracer for the magnetic field direction, especially in the inner regions of the bowshock;
\item Thermal emission can 
outshine
synchrotron emission;
\item The unpolarized thermal emission can greatly reduce the polarization degree;
\item Thermal emission contamination produces nonuniform spectral-index maps as observed;
\item Bowshocks that have been detected at low radio frequencies are { particle accelerating stellar bowshock nebulae} ({PASBNe}) candidates;  
\item An inhomogeneous component of the magnetic field can decrease dramatically the polarization degree at radio frequencies for a ratio $\delta B/B_{\rm reg} = 0.5$ (and higher), and under the assumptions we made in this study, we predict a maximum degree of polarization  of $\approx$ { 40\% (for $\mathbf{\nu = 1.4}$ GHz )} and a most feasible value of { 8-10\%} within the regions of maximum intensity;
\item For most of the inclinations studied here, the region of highest polarization does not coincide with that of the maximum intensity  which directly impacts the detectability;
\item The maximum contribution from the background cosmic rays is lower than the one of the injected electrons. This component adds to the total emissivity, 
but has little effect on the local polarization degree.
\end{enumerate}

\begin{acknowledgments}

M.V.d.V. is supported by the Grants 2019/05757-9 and 2020/08729-3, Fundação
de Amparo à Pesquisa do Estado de São Paulo (FAPESP). M.~V.~d.V. acknowledges support from the Alexander von Humboldt Foundation. The authors thank the anonymous reviewer who helped improve the quality of this work with her/his insightful comments, criticisms and corrections.

\end{acknowledgments}

\bibliographystyle{yahapj}

\bibliography{Yourfile} 

\end{document}